# Raman-scattering-assistant broadband noise-like pulse generation in all-normal-dispersion fiber lasers

**Daojing Li,[1] Deyuan Shen,[1] Lei Li,[2] Hao Chen,[2,*] Dingyuan Tang,[2] and Luming Zhao[2]**

*[1]Department of Optical Science and Engineering, Fudan University, Shanghai 200433, China*
*[2]Jiangsu Key Laboratory of Advanced Laser Materials and Devices, School of Physics and Electronic Engineering, Jiangsu Normal University, Xuzhou, Jiangsu, 221116, China*
*[*]chenhao@jsnu.edu.cn, corresponding author*

We report on the observation of both stable dissipative solitons and noise-like pulses with the presence of strong Raman scattering in a relatively short all-normal-dispersion Yb-doped fiber laser. We show that Raman scattering can be filtered out by intracavity filter. Furthermore, by appropriate intracavity polarization control, the Raman effect can be utilized to generate broadband noise-like pulses (NLPs) with bandwidth up to 61.4 nm. To the best of our knowledge, this is the broadest NLP achieved in all-normal-dispersion fiber lasers.

## 1. Introduction

Passively mode-locked fiber lasers have been extensively investigated in the past two decades as an excellent ultrashort pulse source [1-5]. Dissipative soliton (DS) [1,2] that formed in the all-normal-dispersion fiber laser, attracts considerable attention as it favors larger pulse energy compared to the soliton formed in the anomalous-dispersion regime [3] and the dispersion-managed regime [4]. The pulse is continuously stretched, accumulating large energy and chirp when it propagates in an all-normal-dispersion cavity. Spectral filtering is needed to convert frequency chirp to self-amplitude modulation. Further scaling of the DS pulse energy can be achieved by increasing cavity length or mode-field diameter of the fiber [5].

As the laser pulse energy keeps increasing, high-order nonlinear effects begin to impact the laser performance, especially for powerful DS pulses in cavities with large nonlinearity, where the Raman scattering of the DS imposes another limitation to energy scalability [6-9]. With implementation of a long polarization maintaining (PM) fiber, Kharenko et al. observed and well-studied noisy Raman pulse limiting the achievable DS power in an all-fiber (30-m long) mode locked laser [6-8]. In an all-PM fiber laser configuration, destabilization of the DSs into noise-like pulses (NLPs) was reported for cavity length longer than 100 m, which was attributed to the stimulated Raman scattering [9].

First demonstrated by Horowitz et al. [10], NLP represents a bunch of ultrashort pulses, which has been studied in various cavity configurations. Most of these studies were performed based on the Er-doped fiber (EDF) laser cavities [10-14]. It was shown that NLPs can form very broad output spectra, even broader than the gain bandwidth. The broadest NLP obtained in EDF was 135 nm by Vazquez-Zuniga et al. [14]. NLPs in YDF lasers were less studied [15-17]. Zaytsev et al. reported the generation of NLP with spectral bandwidth up to 48.2 nm from an Yb-doped dispersion-managed (DM) fiber laser [16]. Very recently, Suzuki et al. demonstrated 131 nm NLP in the YDF cavity with cavity dispersion carefully managed near to zero [17]. Broadband NLP generation reported were mainly performed in dispersion managed regime. Pulses are stretched and compressed periodically in DM cavities. The broad spectra are primarily attributed to the enhanced self-phase modulation of the high peak power during the compression inside the cavity. Whereas in all-normal dispersion region, where pulse shaping is based on the joint action of dispersion, nonlinearity and dissipation, pulses are continuously stretched. The peak power is relatively low. No work has been reported on the broadband NLP in all-normal dispersion Yb fiber lasers.

In this Letter, we report on both stable DS and NLP operation with the presence of strong Raman scattering in an all-normal-dispersion YDF DS laser mode locked by the nonlinear polarization rotation technique. In [6-8], the Raman effect came into play at cavity length larger than 30 m. With longer cavity length and higher pump, DS remains stable from the Raman wave. In [9] DS was destabilized into NLP by the Raman scattering when the cavity length is up to 200 m. Different from the previous works [6-9], where the laser cavities were constructed mainly with PM fiber and the cavity length extended from 30 to 200 m, in this work all the fibers used were the standard single-mode fiber (SMF) except the YDF and the total cavity length was much shorter (~11.7 m). It is well-known that Raman effect appears easily for long fiber cavity or for fibers with small mode field diameter. However, both stable DSs and NLPs with strong Raman conversion were obtained, although the cavity was relatively short and consisted of no PM or high nonlinearity components. Carefully adjusting polarization controllers, the Raman conversion can be either successfully filtered out by the birefringence filter, or assistant to the generation of broadband NLP with spectral bandwidth up to 61.4 nm. To the best of our knowledge, this is the broadest NLP obtained in all-normal-dispersion fiber lasers.

## 2. Experimental setup

Figure 1 shows the laser configuration. As mentioned, the cavity was constructed only by non-PM components. A segment of 35 cm YDF (YB406, CorActive) was used as the gain medium. All the other fibers used were the standard SMF (1060XP). The laser was pumped by a 976 nm pump laser through a wavelength division multiplexer (WDM). Two quarter-wave plates and one half-wave plate together with one polarization beam splitter (PBS) were mounted on a 135-mm-long fiber bench to achieve the NPR mode-locking. Notice that this cavity was bidirectional since there was no isolator in the cavity. A 10% fiber coupler was used to monitor intracavity laser operation. No dispersion compensation was deployed in the cavity. The total cavity length was about 11.7 m, which was much shorter than that in previous work (30-200 m).

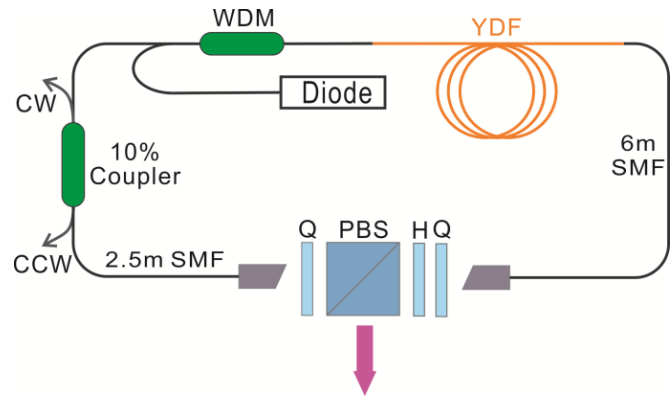

Fig. 1. Schematic of the fiber laser. YDF: Yb-doped fiber; SMF: single mode fiber; WDM: wavelength division multiplexer; PBS: polarization beam splitter; Q: quarter-wave plate; H: half-wave plate; CW: clockwise; CCW: counterclockwise.

## 3. Results and discussions

### 3.1 Dissipative soliton with Raman scattering

With appropriate settings of the plates, self-starting mode locking can be achieved when the pump reaches 496 mW. Once mode locked, the laser operated in the clockwise (CW) direction due to different nonlinearity experienced by pulses along the different directions. Specifically, in the CW direction after being amplified in the YDF, the pulse propagates along a 6 m single-mode fiber (SMF) then arrives at the PBS, while in the count-clockwise (CCW) direction the amplified pulse experiences the cavity output loss before the PBS, accumulating less nonlinear phase shift than the CW pulse. Figure 2(a) shows the spectra of the output pulses from the PBS reflection and the intracavity output from the 10% coupler at 525 mW pump. Although it was a relatively short cavity, a Raman spectrum was detected in addition to the steep edge DS spectrum centered in 1052 nm. The 10% output was 28 mW, corresponding to 1.6 nJ, 1.5% of which belongs to the Raman pulse. The Raman pulse was amplified during propagation along the cavity. Of the PBS output, 10% of the energy belongs to the Raman pulse. The total output power was 60 mW, which corresponds to a pulse energy of 3.4 nJ at 17.82 MHz repetition rate.

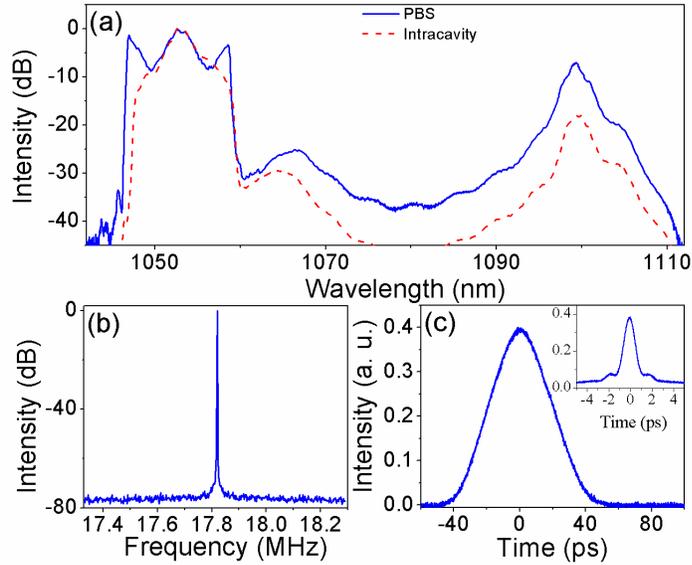

Fig. 2. (a). Normalized optical spectra (solid blue line, the PBS output; dashed red line, intracavity pulse from 10% coupler). (b). RF spectrum of the PBS output. (c). Chirped and dechirped (inset) PBS output pulse autocorrelation traces.

Despite of the strong Raman scattering, the stable DS mode-locking was not affected. The radio frequency (RF) spectrum of the PBS reflection output exhibited high contrast near 80 dB as shown in Fig. 2(b), indicating low-amplitude fluctuations. The resolution bandwidth (RBW) of the RF spectrum analyzer is 10 Hz. The triangle shape autocorrelation trace of the PBS output as shown in Fig. 2(c) implied a pulse duration of 30.4 ps. The pulse can be compressed by a grating pair down to 950 fs (a time-bandwidth product of 1.93). Whereas the 10% output can be dechirped to 750 fs, resulting in a time-bandwidth product of 0.56.

### 3.2 Dissipative soliton without Raman scattering

Since no concrete bandpass filter was used in the cavity, the spectral filtering was assumed to be provided by the intrinsic birefringence filter of the NPR mode-locking [18]. This was proved by the central wavelength shift with different wave plate settings, as shown in Fig. 3. Trace i represents the output state from the PBS port in Fig. 2. In trace j and k, one of the quarter plates was carefully rotated from the original state, and the central wavelength was shifted from 1052.6 nm to 1057.6 nm and 1062.5 nm under fixed pump power of 525 mW. As a result, the wavelength of the Raman spectrum was also shifted correspondingly. Apart from the wavelength movement, another feature was observed: The relative strength of the Raman part decreased. In trace k, the Raman spectral intensity was suppressed down to over 30 dB lower than that of the main DS. The measured pulse duration increased to 36.5 ps. The broader pulse was well explained because all the laser power was concentrated on DSs. Broader pulse duration is expected for larger pulse energy.

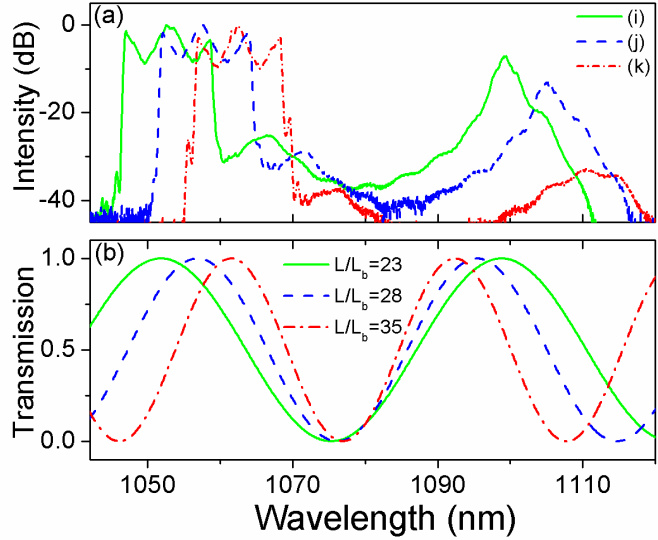

Fig. 3. (a) Variation of the normalized optical spectral with different cavity birefringence under fixed pump power. (b). Cavity transmission versus wavelength under different cavity birefringence. The central wavelength is at 1080 nm, and the cavity linear phase delay bias is $0.8\pi$.

We attribute this suppression of the Raman scattering to the intracavity birefringence filter. DS generation in YDF laser with this birefringence filter has been explicitly studied in previous work [18]. It is well known that the birefringence filter has a sinusoidal transmission function of the linear and nonlinear phase shift of light in the cavity. Varying the cavity birefringence will not only shift the central wavelength, but also affect the transmission bandwidths as well as the period of the sinusoidal filter. To highlight it, we draw in Fig. 3(b) the cavity transmission around 1080 nm under different cavity birefringence. Rotating the wave plate, which is equivalent to changing the cavity birefringence also changes the period of the transmission. In state *i,* the DS and the Raman signal both experienced large transmission. The estimated cavity energy after the YDF was approximately 10 times of the 10% output energy plus the PBS output energy, i.e., 19.4 nJ. Given the high pump power, DSs with strong Raman conversion could be generated. While rotating the wave plate, the Stoke wavelength mismatches with the filter pass-band and its transmission kept decreasing. As a result, in state *k*, the Raman conversion was successfully filtered out by the cavity filter.

### 3.3 Broadbandy noise like pulse induced by Raman scattering

Apart from stable mode-locking operation, NLP emission could also be achieved when the wave plates were significantly tuned away from the mode-locking conditions. Under same pump power, by carefully adjusting the wave plates, along with the cavity filter, the spectral of the main NLP pulse and the Raman conversion part could connect together. Consequently, a broadband NLP was generated in the cavity. Its spectrum was shown as trace m in Fig. 4(a). Under NLP operation, increased pump power tended to fill up the gaps between the main pulse and Stoke signal. At the maximum pump power of 1150 mW, broad flat spectrum with up to 61.4 nm bandwidth was obtained at the PBS port [trace n in Fig. 4(a)]. The output power was 195 mW, corresponding to 11 nJ. The autocorrelation trace was shown in Fig. 4(c). The grooves at the right side of autocorrelation trace were due to our autocorrelator itself. The trace exhibited a 131 fs coherent spike on the top of a broad pedestal, which was a typical autocorrelation profile of NLPs. The RF spectrum exhibited high contrast near 70 dB (with RBW of 30 Hz). This is, to the best of our knowledge, the broadest NLP generation in all-normal-dispersion fiber lasers.

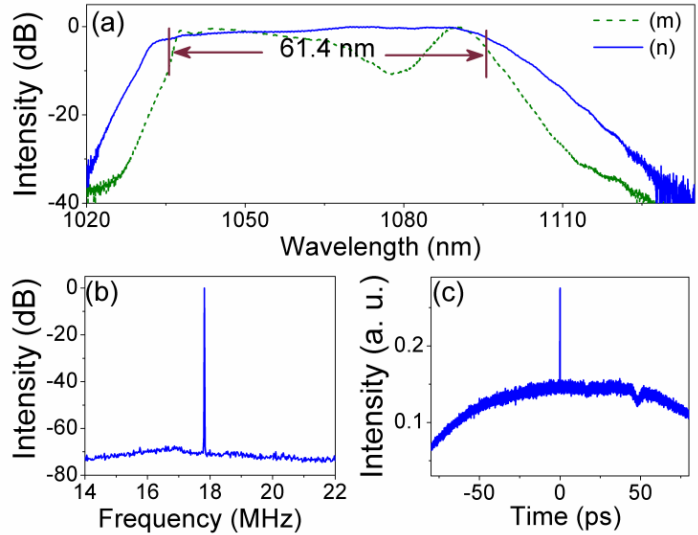

Fig. 4. Broadband NLP from the PBS reflection. (a) Optical spectra (b) RF spectrum and (c) Autocorrelation trace of the NLP at pump power of 525 mW (m-dashed green line) and 1150mW (n-solid blue line).

In the Raman process, a Stokes wave is induced by the DS. Therefore, the Stokes wave is generated from and amplified by the DS pulse. The DS and Stokes pulses have different group velocity because of dispersion. Hence, the generated Stokes pulses will quickly separate away from the DS. And the Stokes pulses start to attenuate, and finally collapse. With high cavity energy, Stokes pulses are constantly generated and decayed. In the normal dispersion regime, due to the small group velocity difference and the short propagation length, the Raman pulses and the DS manage to travel together in the cavity. It forms the experimentally observed ultrafast bundles. It is reasonable to consider it as a broadband noise-like pulse instead of a NLP together with a separated Raman wave. We note that work on broadband NLP have utilized Raman effect to broaden spectra [13,14].

## 4. Conclusion

In conclusion, we have experimentally demonstrated both stable DS and NLP operation with the presence of strong Raman scattering in an all-normal-dispersion YDF laser mode-locked by NPR. We note that the cavity was relatively short and no PM or high nonlinear component was included. NPR mode-locking technique generates periodical transmission filter in the cavity. When the Raman wavelength shift matches with the period of sinusoidal filter, that is, the Raman conversion also experienced large transmission, stable DSs with strong Raman conversion were generated with high pump power. We show that the Raman conversion could be successfully filtered out by the intracavity filter. As a result, introducing a filter with single pass-band is suggested to circumvent the Raman conversion in practice for purchasing high energy DSs. On the other hand, Raman scattering could be used to generate broadband NLP. In this work, a NLP with bandwidth up to 61.4 nm was achieved, which is the broadest NLPs obtained in all-normal-dispersion fiber lasers.

### Acknowledgement

This work was supported in part by the National Natural Science Foundation of China under Grant 61177045, 11274144, 61275109 and 61405079, in part by the Priority Academic Program Development of Jiangsu higher education institutions (PAPD), in part by the the Jiangsu Province Science Foundation (BK20140231).